\documentclass[a4paper]{article}
\usepackage[psamsfonts]{amssymb}
\usepackage{amsmath}
\usepackage{epsfig}
\title{ \bf Plane symmetric solutions in Ho$\check{\textbf{r}}$ava-Lifshitz theory}
\author{\normalsize{M. R. Setare\thanks{%
E-mail: rezakord@ipm.ir}  \, and \,D. Momeni  \thanks{%
E-mail: dmomeni@phymail.ut.ac.ir} }\\
\newline
\\
{\normalsize \it  Department of Science of Bijar, University of
Kurdistan, Bijar, Iran} }
\date{}

\begin{document}
\maketitle
\newcommand{\be}{\begin{equation}}
\newcommand{\ee}{\end{equation}}
\newcommand{\bq}{\begin{eqnarray}}
\newcommand{\eq}{\end{eqnarray}}
\vspace{1cm}
\begin{abstract}
 The purpose of this paper is to find and analyze plane symmetric, static(non static)
  solutions in Ho\v{r}ava- Lifshitz gravity. We discussed two versions of Horava
gravity. First we showed that if the detailed balance principle
have considered, there are both static and non-static solutions.
We show that in static case there are two family of solvable
models which either of them has a well defined EOS, in analogous
to the perfect fluid solutions in GR. In non-static case we find
a family of solutions. Some physical properties of these
solutions was discussed. Secondly we investigated the plane
symmetric solutions for a new modified version of Ho\v{r}avaa
gravity \cite{bla}, which has the new terms inserted action in it.

\end{abstract}

\section{Introduction}
The Einstein's equations for hypersurface-homogeneous space-time are
reduced to a system of ordinary differential equations. At least
for the spatially-homogeneous case, they form a well-posed Cauchy
problem \cite{1} and, although they have not been completely
integrated, their qualitative properties have been discussed in
many papers: (see Wainwright and Ellis \cite{2} for an extensive
survey of the results, and e.g.\cite{3} Ryan and Shepley ,
MacCallum \cite{4}, Bogoyavlenskii \cite{5}, and Rosquist and
Jantzen \cite{6} for useful earlier reviews). Methods of
dynamical systems theory which proved fruitful in elucidating
these properties lead to ways of restricting the general
case to more readily solvable subcases and thence to new exact
solutions (see e.g. Uggla et al. \cite{7} for a summary). Nearly
all of these methods were initially developed  for use in the
spatially-homogeneous case ('cosmologies', for brevity), and we
shall therefore describe the methods in this context although
they can be adapted to the $G_{3}$ on $T_{3}$ and $H_{3}$ on
$V_{3}$ cases also (e.g. for an orthonormal tetrad method for G3
on T3 see Harness \cite{8}). The number of  freedom degrees ,
i.e. the number of essential arbitrary constants required in a
general cosmology for each Bianchi type, has been studied by
Siklos \cite{9} and MacCallum \cite{10}, Wainwright and Ellis
\cite{11}). The residual set of ordinary differential equations which
are going to be solved can be formulated in various ways \cite{4}, Wainwright
and Ellis \cite{2}. One can use a time-independent basis as in
and parameterize the components $g_{\alpha\beta}$ in
suitable ways. This is called the metric approach. One may then
choose spatial
coordinations and a new time coordination $\tau$.\\
The main alternative to the metric approach is the ortho normal
tetrad approach using a tetrad basis. The two  ways are closely related
when variables are chosen as just described above \cite{11},
in both approaches a suitable choice of lapse (or, to include the
$G_{3}$ on $T_{3}$ case, 'slicing gauge'), or directly of
independent variable, may decouple and simplify the equations; for
details see e.g. Jantzen \cite{12}, Uggla et al. \cite{7}. A
power-law lapse, i.e. a product of powers of the dependent
variables, an idea introduced in Bonanos \cite{13}, may be useful,
as may an 'intrinsic slicing' (e.g. making a product of
invariantly-defined metric components the independent variable).\\
Recently, a power-counting renormalizable, ultra-violet (UV)
complete theory of gravity was proposed by Ho\v{r}ava in
\cite{hor2,hor1,hor3,hor4}. Although presenting an infrared (IR)
fixed point, namely General Relativity, in the  UV the theory
possesses a fixed point with an anisotropic, Lifshitz scaling
between time and space of the form $x^{i}\to\ell~x^{i}$,
$t\to\ell^z~t$, where $\ell$, $z$, $x^{i}$ and $t$ are the scaling
factor, dynamical critical exponent, spatial coordination and
temporal coordination, respectively. But recently Blas, et.al listed
some inconsistencies of the Ho\v{r}ava-Lifshitz gravity as a
complete description of Quantum gravity. They discussed  the
consistency of Ho\v{r}ava's proposal for a theory of quantum
gravity from the low-energy viewpoint. They uncover the
additional scalar freedom degree arising from the explicit
breaks the general covariance and its followed properties.
Their analysis was done both in the original formulation of the
theory and in the St$¨$uckelberg frame. A unusual feature of the
new mode is that it satisfies a first order(in time derivatives)
equation of motion . In linear aproximation this extra freedom degree
manifested only around non-static spatially inhomogeneous
backgrounds. They found two serious problems associated with this
mode. "\emph{First, the mode develops very fast exponential
instabilities at short distances. Second, it becomes strongly
coupled at an extremely low cutoff scale}". They also discussed a
version of Ho\v{r}ava's gravity with projectable condition and
stated that this version is a certain limit of the ghost
condensate model. The theory is still problematic since the
additional field generically forms caustics and, again, has a
very low strong coupling scale. Also they clarify some subtleties
that arise from the application of the St$¨$uckelberg formalism to
Ho\v{r}ava's model due to its non-relativistic nature.\\
 Due to these features, there has been a large effort on
examining and extending the properties of the theory itself
\cite{Volovik:2009av,Orlando:2009en,Nishioka:2009iq,Konoplya:2009ig,Charmousis:2009tc,
Li:2009bg,Visser:2009fg,Sotiriou:2009bx,Chen:2009bu,Chen:2009ka,Shu:2009gc,Bogdanos:2009uj,Kluson:2009rk,
chen}. Additionally, application of Ho\v{r}ava-Lifshitz gravity
as a cosmological framework gives rise to Ho\v{r}ava-Lifshitz
cosmology, which is going to lead from interesting behavior
\cite{Calcagni:2009ar,Kiritsis:2009sh, odi, sari1, sari2}. In
particular, one can examine specific solution subclasses
\cite{Lu:2009em,Nastase:2009nk,Minamitsuji:2009ii}, the
perturbation spectrum \cite{cai,
Gao:2009bx,Chen:2009jr,Gao:2009ht,Wang:2009yz,Kobayashi:2009hh,
ding}, the gravitational wave production
\cite{Takahashi:2009wc,Koh:2009cy}, the matter bounce
\cite{Brandenberger:2009yt,Brandenberger:2009ic,Cai:2009in}, the
black hole properties \cite{Danielsson:2009gi,Kehagias:2009is,
Mann:2009yx,Bertoldi:2009vn,Castillo:2009ci,BottaCantcheff:2009mp,
ohta:cao}, the cosmic string solutions \cite{mom} the dark energy
phenomenology
\cite{Saridakis:2009bv,Wang:2009rw,Appignani:2009dy,Setare:2009vm},
the astrophysical phenomenology \cite{Kim:2009dq,Harko:2009qr}
etc. However, despite this extended research, there are still many
ambiguities if Ho\v{r}ava-Lifshitz gravity is reliable and capable
from a successful description for the gravitational background of our
world, there still many ambiguities.
\cite{Charmousis:2009tc,Sotiriou:2009bx,Bogdanos:2009uj}.\\
In the present paper we would like to obtain plane-symmetric
perfect fluid solutions in the framework of Ho\v{r}ava-Lifshitz
gravity. At first we review such solutions in both static and
non-static cases in the framework of general relativity. After a
brief review of Ho\v{r}ava-Lifshitz gravity, we obtain and
analyze plane symmetric, static Taub-like solution. We extend our
study to the Purely time dependent solution.Also we discussed
some simple solutions for a new version of modified Horava theory
which was recently proposed by Blas , et al.

\section{Plane-symmetric perfect fluid solutions in GR}
\subsection{ Static solutions}

The plane-symmetric static perfect fluids with a prescribed
equation of state $\mu=\mu(p)$ are given by \cite{14,15}
\begin{eqnarray}
ds^2=-e^{2\nu}dt^2+z^{2}(dx^2+dy^2)+zF(z)^{-1}dz^2
\end{eqnarray}
where

\begin{eqnarray}
\frac{2zp'}{\mu(p)+p}=1-\kappa_{0}p\frac{z^3}{F}=-2z\nu'\\
F'=-\kappa_{0}\mu(p)z^2
\end{eqnarray}
For a given function $\mu=\mu(p)$, the differential equations
 determine F(z) and $p=p(z)$ and from
p(z) then $\nu=\nu(z)$. Equations (2,3) lead to the
condition\cite{16}

\begin{eqnarray}\nonumber
\frac{p'}{p}=\frac{z^2+G'}{G}\frac{G+z^3}{z^3-G}\\\nonumber
G=-\frac{F(z)}{\kappa_{0} p(z)}
\end{eqnarray}
So one may prescribe G(z) and then obtain p and $\nu=\nu(z)$ as
linear integrals. For an equation of state $p = (\gamma-1)\mu$ the
function $G$ was found by Collins \cite{17}. A solution for $p =
\mu/3$ was obtained by Teixeira et al. \cite{18}. The solution
for $\mu = const$ was given by Taub \cite{14} and by Horsk´y
\cite{19} in terms of hypergeometric functions. Some other
special solutions have been given by Davidson \cite{20}. Static
plane-symmetric perfect fluids occur also as subcases of the
static cylindrically-symmetric solutions.

\subsection{ Non-static solutions}
Besides the classes of solutions which were described in $a$, several other
classes have been found by making special assumptions for the
metric or the equation of state. For $p = \mu$, Tabensky and Taub
\cite{15} reduced the field equation to a single linear
differential equation,
\begin{eqnarray}
ds^2 = t^{-1/2}e^{\Omega}(dz^2 - dt^2) + t(dx^2 +dy^2), t>0\\
\Omega= 2\int
t[(\sigma_{,t}^2+\sigma_{,z}^2)dt+2\sigma_{,t}\sigma_{,z}
dz]\\\sigma_{,tt} + t^{-1}\sigma_{,t}-\sigma_{,zz} =
0\\\kappa_{0}p = \kappa_{0}\mu =
t^{1/2}e^{-\Omega}(\sigma_{,t}^2-\sigma_{,z}^2),
\end{eqnarray}
Tabensky and Taub \cite{15} also gave the special solution
\footnote{$\log(x)=\int^{x}_{1}\frac{dt}{t}$}
\begin{eqnarray}
\sigma=\alpha \log(t) +\beta \arccos(\frac{z}{t})\\
\Omega=2(\alpha^2+\beta^2)\log(t)+2\beta^2\log(1-\frac{z^2}{t^2})+4\alpha\beta
\arccos(\frac{z}{t})
\end{eqnarray}
This solution represents the asymptotic Robertson-Walker
Space-time. These functions are  particular solutions which are
inhomogeneous and anisotropic but toward to an homogeneous
spacetime for large times. The solution depends on two parameters
$\alpha$ and $\beta$. For $\alpha^2+\beta^2=\frac{3}{2}$. We get
Robertson- Walker space-time asymptotically.The variables $z$ and
$t$ are restricted to $z^2\leq t^2$.

For further solutions see Collins and Lang \cite{21}, Goode
\cite{22}, Carot and Sintes \cite{23}, Bray \cite{24}, Tariq and
Tupper \cite{25}, Shikin \cite{26}, Gupta and Sharma \cite{27}. We
want to obtain the analogous of the metric solution (3) in Horava
theory. We use from this gauge, since firstly it has a well known
simple GR analogous and secondly since the field equations has
only one gauge function $\Omega$ and also this solution is
\emph{spatially Ricci flat}, i.e $R=0$ and several terms in Horava
action will be vanished, especially the \emph{Cotton }tensor $C_{ij}=0$.

\section{Review of Ho$\check{\textbf{r}}$ava-Lifshitz gravity with detailed condition}

 Following from the ADM decomposition of the metric  \cite{28},
  and the Einstein equations, the fundamental objects of interest
are the fields $N(t,x),N_{i}(t,x),g_{ij}(t,x)$ corresponding to
the \emph{lapse }, \emph{shift} and \emph{spatial metric} of the
ADM decomposition.
 In the $(3 + 1)$-dimensional ADM formalism, where the
metric can be written as
 \begin{eqnarray}
ds^2=-N^2dt^2+g_{ij}(dx^{i}+N^{i}dt)(dx^{j}+N^{j}dt)
\end{eqnarray}
and for a spacelike hypersurface with a fixed time, its extrinsic
curvature $K_{ij}$ is

\begin{eqnarray}\nonumber
K_{ij}=\frac{1}{2N}(\dot{g_{ij}}-\nabla_{i}N_{j}-\nabla_{j}N_{i})
\end{eqnarray}
where "dot" denotes a derivative from "t" and covariant
derivatives defined with respect to the spatial metric $g_{ij}$,
the action of Ho$\check{\textbf{r}}$ava-Lifshitz theory  for
$z=3$ is
\begin{eqnarray}
S=\int_{M} dtd^{3}x\sqrt{g} N(\mathcal{L}_{K} - \mathcal{L}_{V} )
\end{eqnarray}
we define the space-covariant derivative on a covector $v_{i}$ as
$\nabla_{i}v_{j}\equiv \partial_{i}v_{j}-\Gamma_{ij}^{l}v_{l}$
where $\Gamma_{ij}^{l}$ is the spatial Christoffel symbol.$g$ is
the determinant of the 3-metric and $N = N(t)$ is a dimensionless
homogeneous gauge field. The kinetic term is
\begin{eqnarray}\nonumber
\mathcal{L}_{K}=\frac{2}{\kappa^2}\mathcal{O}_{K}=\frac{2}{\kappa^2}(K_{ij}K^{ij}-\lambda
K^2)
\end{eqnarray}
Here $N_{i} $ is a gauge field with scaling dimension $[N_{i}] =
z -
1$.\\
The \emph{'potential'} term $\mathcal{L}_{V}$ of the
$(3+1)$-dimensional theory is determined by the \emph{principle of
detailed balance }\cite{3}, requiring $\mathcal{L}_{V}$ to follow,
in a precise way, from the gradient flow generated by a
3-dimensional action $W_{g}$. This principle was applied to
gravity with the result that the number of possible terms in
$\mathcal{L}_{V} $ are drastically reduced with respect to the
broad choice available in an '\emph{potential} is
\begin{eqnarray}
\mathcal{L}_{V}=\alpha_{6}C_{ij}C^{ij} -
\alpha_{5}\epsilon_{l}^{ij} R_{im}\nabla_{j}R^{ml} + \alpha_{4}
[R_{ij}R^{ij}- \frac{4\lambda-1}{4(3\lambda-1)} R^2]
+\alpha_{2}(R - 3\Lambda_{W})
\end{eqnarray}
The coupling constants $\alpha_{i}$ define by
\begin{eqnarray}\nonumber
\alpha_{2}=\frac{\alpha_{4}\Lambda_{w}}{3\lambda-1}\\\nonumber
\alpha_{4}=\frac{\kappa^2\mu^2}{8}\\\nonumber
\alpha_{6}=\frac{\kappa^2}{2\nu^4}\\\nonumber\alpha_{5}=\frac{\kappa^2\mu}{2\nu^2}
\end{eqnarray}

 Where in it $C_{ij}$
is the \emph{Cotton }tensor \cite{3} which is defined as,
\begin{eqnarray}\nonumber
C^{ij}=\epsilon^{kl(i}\nabla_{k}R^{j)}_{l}
\end{eqnarray}
Following \cite{Lu:2009em} we can write the action as
\begin{eqnarray}
S=\int dtdx^3(\mathcal{L}_{0}+\mathcal{L}_{1})\\
\mathcal{L}_{0}=\sqrt{g}N(\frac{2}{\kappa^2}(K_{ij}K^{ij}-\lambda
K^2)+\frac{\kappa^2\mu^2(\Lambda_{w}R-3\Lambda_{w}^2)}{8(1-3\lambda)})\\
\mathcal{L}_{1}=\sqrt{g}N(\frac{\kappa^2\mu^2(1-4\lambda)}{32(1-3\lambda)}R^2-\frac{\kappa^2}{2w^4}(C^{ij}-\frac{\mu
w^2}{2}R^{ij})(C_{ij}-\frac{\mu w^2}{2}R_{ij}))
\end{eqnarray}

\section{Exact solutions}
We shall restrict ourselves to situations where the space time has
plane symmetry.the problem is invariant under
transformations of the form
\begin{eqnarray}\nonumber
x\rightarrow x+a\\\nonumber y\rightarrow y+b\\\nonumber
,\\\nonumber x\rightarrow x\cos(\theta)+y\sin(\theta)\\\nonumber
y\rightarrow y\cos(\theta)-x\sin(\theta)
\end{eqnarray}
The metric of a spacetime which admits plane symmetry may be written
as \cite{28}
\begin{eqnarray}
ds^2=-e^{2F}dt^2+\frac{1}{c^2}(e^{2H}(dx^2+dy^2)+e^{2G}dz^2)
\end{eqnarray}
Where $F,G$ and $H$ are just functions  of $z$ and $t$ alone.
By substituting (16) in (14,15) finally we obtain the following form
of action
\begin{eqnarray}
S=\int dt\int
d^3x[\frac{1}{c^3}e^{2H+G+F}[\frac{2}{\kappa^2}e^{-2F}[(1-\lambda)\dot{G}^2+
2(1-2\lambda)\dot{H}^2-4\lambda\dot{G}\dot{H}]+\frac{\kappa^2\mu^2}{8(1-3\lambda)}(\Lambda_{w}R-3\Lambda_{w}^2)\\\nonumber+
\frac{\kappa^2\mu^2(1-4\lambda)}{32(1-3\lambda)}R^2-\frac{\kappa^2\mu^2c^4}{8}e^{-4G}((2\acute{H}^2+H''-H'G')^2+(2\acute{H}^2+2H''-2H'G')^2)]]
\end{eqnarray}
Where in it $f'\equiv\frac{df}{dz},\dot{f}\equiv\frac{df}{dt}$ and
the \emph{spatial Ricci Scalar} for metric (15) reads as,
\begin{eqnarray}\nonumber
R=2c^2e^{-2G}(3H'^2+2H''-2H'G')
\end{eqnarray}
 By varying $G,H$
and $F$, we obtain the equations of motion. Considering the case
where $3\lambda-1 > 0$, one can see, from the modified Friedmann
equations in the context of the Horava theory, a negative
$\Lambda$ is required. In Ref. \cite{Lu:2009em}, the authors point
out that a positive $\Lambda$ can be obtained by making an
analytical continuation for parameters. In this work we emphasized
on the$3\lambda-1 \leq 0$. Choosing $\lambda$ like above avoids us
from a negative cosmological constant term which is so harmful
for the cosmological evidences.

\subsection{Static Taub-like solution}
In this section we want to obtain a Taub like solution (1) as
metric functions (2,3) with action (17). Comparing two metrics
(16) and (1)we observe that we must take the metric functions as
$H=\log(cz)$,$G=\log(c\sqrt{\frac{z}{F(z)}})$ and $F=\nu(z)$. In
this case the scalar Ricci is $R=2\frac{F'}{z^2}$ and the full
action (17) convert to the next functional:
\begin{eqnarray}
S=\int dt d^3x \Phi(\nu,F(z),F'(z))\\
\Phi(\nu,F(z),F'(z))=-\frac{1}{32}{c}^{3}{e}^{\nu}{\kappa}^{2}{\mu}^{2}{z}^{-3.5}{F(z)}^{-0,5}(4\lambda-1)^{-1}(3\lambda-1)^{-1}\sum_{n=0}f_{n}z^{n}
\end{eqnarray}
The $\nu$ equation of motion is:
\begin{eqnarray}
\sum_{n=0}f_{n}z^{n}=0
\end{eqnarray}
Where the Functional coefficients $f_{n}$ were defined by:
\begin{eqnarray}
f_{0}=(24{\lambda}^{2}-11\lambda+1){F}^{2}\\
f_{1}=(-144{\lambda}^{2}+90\lambda-14)FF'\\
f_{2}=(24{\lambda}^{2}-11\lambda+1){F'}^{2}\\
f_{3}=8\Lambda_{w} (4\lambda-1)F\\
f_{4}=8\Lambda _{w}(4\lambda-1)F'\\
f_{6}=12{\Lambda_{w}}^{2}(1-4\lambda)
\end{eqnarray}
Just $F(z)$ field equation is more complicated, we do not write it
here.We mentioned here that the general solution for metric
function $F(z)$ and the energy density function $\rho$ for action
(19) is:
\begin{eqnarray}\\\nonumber
 F(z)=-e^{-1}I(a,b;z)[12\lambda(4\Lambda-1)I(c,d;z)+(1-3\lambda)c_{1}]\\\nonumber
h=8\lambda(z-3)-z+7\\\nonumber I(\alpha,\beta;z)=\int
h^{\alpha}z^{\beta}dz\\\nonumber
\\\kappa_{0}\rho=e^{-1}z^{-2}[h^{a}z^{b}(12\lambda(4\Lambda-1)I(c,d;z)+(1-3\lambda)c_{1})+12\Lambda(4\lambda-1)h^{c}z^{d}I(a,b;z)]
\end{eqnarray}
Where
\begin{eqnarray}\nonumber
 a=-\frac {704\lambda^{2}-400\lambda+55}{(
8\lambda-1 )  ( 24\lambda-7 ) }\\\nonumber b=-8\frac {4
\lambda-1}{24\lambda-7}\\\nonumber c=16\frac { ( 4\lambda-1 )
 ( 8\lambda-3 ) }{ ( 8\lambda-1 ) ( 24
\lambda-7 ) }\\\nonumber d=2\frac
{40\lambda-11}{24\lambda-7}\\\nonumber e= (-1+3\lambda ) ^{-1}
\end{eqnarray}
the integral can be calculated by using the Hypergeometric functions
as\footnote{$F[a,b,c;x]=\sum_{k=0}\frac{(a)_{k}b_{k}x^{k}}{c_{k}k!},
(a)_{k}=\frac{(a+k-1)!}{(a-1)!}$}
\begin{eqnarray}
I(c,d;z)=\int h^{c}z^{d}
dz=\frac{z^{1+d}}{1+d}(7-24\lambda)^{c}F[1+d,-c,2+d,\frac{z(-8\lambda+1)}{7-24\lambda}]
\end{eqnarray}

 Without overloading the paper with lengthy formulae, the
pressure $p(z)$ and metric function $\nu(z)$ can be calculated
from (2,3). It was argued by Horava that in the infrared limit the
higher order terms become negligible and so one may expect to
recover GR if the parameter $\lambda$ flows to 1 in that limit.
But Blas et al \cite{bla} argued that this expectation is
incorrect: the explicit breaking of general covariance leads to
the appearance of an extra degree of freedom in the infrared which
becomes strongly coupled when $\lambda$ approaches 1. More
precisely, the additional mode is weakly coupled only in a narrow
window at low energies. This window depends both on $\lambda$ and
the parameters of the background geometry; it shrinks to zero
when $\lambda\rightarrow 1$ or when the background curvature
vanishes. Thus the solution has no familiar GR limit. Also as
stated by Blas et al \cite{bla}, the only consistent solution for
this theory must be time dependent and inhomogeneous. Another
reason for considering the plane symmetric metrics goes to the
Mukohyama work \cite{82}. He considered the version without
detailed balance condition with the projectablility condition and
addressed one aspect of the theory: avoidance of caustics for
constant time hypersurfaces. He showed that \emph{there is no
caustic with plane symmetry in the absence of matter source} if
$\lambda\neq 1$. If $\lambda=1$ is a stable IR fixed point of the
renormalization group flow then $\lambda$ is expected to be deviated
from 1 near would-be caustics, where the extrinsic curvature
increases and high energy corrections become important. Therefore,
the absence of caustics with $\lambda\neq 1$ implies that
caustics cannot form with this symmetry when the source is absent.
He argued that inclusion of matter source will not change
the conclusion. He also argued that caustics with codimension
higher than one will not form because of repulsive gravity
generated by nonlinear higher curvature terms. These arguments
support Mukohyama conjecture that there is no caustic for
constant time hypersurfaces. Our exact plane symmetric solution
endorsement this conjecture directly.

\subsection{Other special solutions}
\textbf{Case} $\lambda=1/3$:\\
As the observation in \cite{hor3} the value $\lambda=1/3$
corresponds to the action being invariant under an anisotropic
conformal (Weyl) symmetry. The case $\lambda=1/3$, corresponds
to an anisotropic Weyl-invariant theory, is another relevant
value for which the equations of motion simplify and hence admit
quite explicit solutions. Recently \cite{84} derived general
static spherically symmetric solutions in the Ho\v{r}ava theory of
gravity with nonzero shift field. These represent "hedgehog"
versions of black holes with radial "hair" arising from the shift
field. For the case of the standard de Witt kinetic term
($\lambda=1$) there is an infinity of solutions that exhibit a
deformed version of re parametrization invariance away from the
general relativistic limit. Special solutions also arise in the
anisotropic conformal point $\lambda=1/3$. This is a simple reason
that why we
investigated solution with $\lambda=1/3$.\\
 The  field equation is:
\begin{eqnarray}
\Lambda_{w} {z}^{3}=\frac{2}{3}(z F)'
\end{eqnarray}
The general solution for this differential  equation is:
\begin{eqnarray}
F(z)=\frac{3\Lambda_{w}}{8}{z}^{3}+c_{1}/z
\end{eqnarray}
To find the unknown metric function $\nu$ we must use
(2,3). By solving  the simple differential equations for
$\nu(z),p(z)$ and by substituting $\mu(z)$ from (3), after a
simple integration it leads to the\footnote{ $P(a,b,x)$, $Q(a,b,x)$
are the associated \emph{Legendre} functions of the first and
second kind, new coordinate $\xi$ is defined as
$\xi=(3\Lambda_{w} z^{4}+8 c_{1})^{1/2} $}:
\begin{eqnarray}
\mu(z)=1/8(-\frac{9\Lambda_{w}}{\kappa_{0}}+8\frac{c_{1}}{\kappa_{0}
{z}^{4}})\\
p(z)=-0.3735\xi\Lambda\frac{\theta}{\eta}\\ \theta=\sqrt{c_{1}}\xi
P(-0.25,1.0308,0.3535\frac{\xi}{\sqrt{c_{1}}})-3.1768
c_{1}P(0.75,1.0308,0.3535\frac{\xi}{\sqrt{c_{1}}})\\\nonumber
+c_{2}\sqrt{c_{1}}\xi
Q(-0.25,1.0308,0.3535\frac{\xi}{\sqrt{c_{1}}})-3.1768c_{1}c_{2}Q(0.75,1.0308,0.3535\frac{\xi}{\sqrt{c_{1}}})\\
\eta=\sqrt{c_{1}}\kappa_{0}(-\xi^2+8c_{1})(c_{2}Q(-0.25,1.0308,0.3535\frac{\xi}{\sqrt{c_{1}}})+P(-0.25,1.0308,0.3535\frac{\xi}{\sqrt{c_{1}}}))
\end{eqnarray}
The expression  of the metric function  $\nu$ is volumeless and
we do not write it here. We must mention here that this spacetime
represents a new exact  solution for GR with a complicated
implicit EOS  $p=p(\mu)$ which could be derived by eliminating
the $z$ coordinate between the both  pressure and energy density
expressions given by the above functions.
\\\textbf{Case} $\lambda=1/4$:\\
In this case the differential equation (20) is:
\begin{eqnarray}
\frac{1}{2}{z}^{2}(F')^{2}+z FF'+\frac{1}{2}{F}^{2}=0
\end{eqnarray}
This equation has a simple solution
\begin{eqnarray}
F(z)=\frac{c_{1}}{z}
\end{eqnarray}
this is the special case of solution (28) with the
$\Lambda_{w}=0$. We can write the following expressions for
$\mu(z),p(z)$:
\begin{eqnarray}
\mu(z)=\frac{c_{1}}{\kappa_{0}{z}^{4}}\\
p(z)=\mu(z)(4-\sqrt{17}\tanh{\frac{\sqrt{17}}{2}(c_{2}-\log{z}}))
\end{eqnarray}
by eliminating the $z$ between (35,36) we can write the EOS as:
\begin{eqnarray}
p(\mu)=\mu(4-\sqrt{17}\tanh{\frac{\sqrt{17}}{2}(c'_{2}+1/4\log{\mu}}))
\end{eqnarray}
Where $c'_{2}=c_{2}-1/4\log(c_{1}/\kappa_{0})$.

 The metric function $\mu(z)$ in this case has simple
expression as:
\begin{eqnarray}
\nu(z)=\log({{z}^{3/2}\cosh{\sqrt{\frac{17}{2}}(\log{z}-c'_{2}}}))
 \end{eqnarray}

Thus the metric (16) transforms to the following form:
\begin{eqnarray}
d{s}^2={z}^{2}[-{z}^{-1/2}\cosh(\sqrt{\frac{17}{2}}(\log{z}-c'_{2}))d{t}^{2}+d{x}^{2}+d{y}^{2}+\frac{1}{c_{1}}d{z}^{2}]
\end{eqnarray}

\subsection{ Purely time dependent solution}
If we consider the non-static, plane symmetric solutions with the
metric anstaz(4).
 In synchronous time $t$, the Plane \emph{ADM } metric
has $N_{i}=0$ , where $\Omega$ is the usual \emph{Tabensky} and
\emph{Taub} metric function. we assumed that the metric function
$\Omega=\Omega(t)$.Since The metric (3) is \emph{Spatial Ricci
flat}, and further the Spatial Ricci tensor and consequently the
result \emph{Cotton }tensor are vanished, then the full curvature
dependence part of the action (13), i.e the
$\mathcal{L}_{1}=0$.The easiest way to obtain the solution for
the full Lagrangian is to substitute the metric anstaz (3) into
the action, and then vary the function $\Omega$. This is a valid
procedure in analogous with cylindrically and spherical kinds.
The result reduces Lagrangian to an overall scaling
constant, is given by
\begin{eqnarray}
\mathcal{L}=\sqrt{t}e^{\Omega}(\frac{2}{\kappa^2}[t^{-3/2}e^{-\Omega}][\frac{1}{2}+
\frac{1}{4}t^{5/2}(-\frac{1}{2}t^{-5/4}+t^{-1/4}\dot{\Omega})^2-\frac{\lambda}{4}(\frac{1}{2}+t\dot{\Omega})^2]-
3\frac{\kappa^2\mu^2}{1-3\lambda}\Lambda_{w}^2)
\end{eqnarray}
Perhaps surprising power $5/2 $ can be understood by observing
that this form of $\mathcal{L}$ is invariant under the reflection.
we assumed that the metric function $\Omega=\Omega(t)$. The
remaining equation is obtained by variation of Lagrangian (15).
The field equation is:
\begin{eqnarray}
\frac{d}{dt}(t\frac{d \Omega}{dt})=b\sqrt{t}e^{\Omega}
\end{eqnarray}
where
\begin{eqnarray}
b=\frac{3\kappa^4\mu^2\Lambda_{w}^2}{(3\lambda-1)(\lambda-1)}
\end{eqnarray}
The general solution for this ODE is
\begin{eqnarray}
\Omega(t)=\log(\frac{(8c_{1}b-9)\sec^{2}(\frac{1}{4}\sqrt{8c_{1}b-9}(c_{2}-\log(t))}{8bt^{3/2}})
\end{eqnarray}
 Thus the general solution for metric (3) with Lagrangian (15) in context of Horava
theory is:
\begin{eqnarray}
ds^2=\frac{(8c_{1}b-9)\sec^{2}(\frac{1}{4}\sqrt{8c_{1}b-9}(c_{2}-\log(t))}{8bt^{2}}(dz^2-dt^2)+t(dx^2+dy^2)
\end{eqnarray}
The interpretation is as follows: we choose $d$ as the proper
distance from $z=0$ along $x, y, t$ constant. The density vanishes
at $z=\infty$ and at $z=0$. The plane $z=0$ behaves like a hard
wall. It can be shown that all time-like geodesics oscillate
between two $z$ planes determined by initial conditions. No
time-like geodesic ever touches the wall. We can,
imagine the space-time for $z<0$ to be the vacuum solution.
Strictly speaking that one doesn't know the matching condition for
singular hypersurfaces but the type of divergence from both sides
of $z=0$ is exactly the same. We conclude that the only way to
have a static solution is to introduce an additional boundary
condition.We shall now show that in asymptotic limit, space-time
is homogeneous, and for some of the parameter values,
Robertson-Walker\footnote{\emph{Asymptotic Robertson-Walker
Space-time}}. All we have to do is to carry out the inversion for
comoving coordinatations for large $t$s . The calculation is
straight-forward but rather long, so we omit it.Therefore as
similar to the GR, we get that for any group of fluid
world-lines, the space-time is homogeneous.If the four-velocity
is irrotational, it can be written in terms of a scalar function
$\sigma(z, t)$ as $u_{n}
=(-{\sigma}^{\alpha}{\sigma}_{\alpha})$,  and \emph{Modified
Einstein's} field equations for a stiff fluid read $p=\rho$.

\section{Plane symmetric solution  for modified Horava-Lifshitz gravity}
In this section we attempt to find analytic solutions for the
modified Horava-Lifshitz gravity proposed recently by Blas, et. al
\cite{bla}. For this we  choose the action to be
\begin{eqnarray}
S=\int d^{3}xdtN\sqrt{g}(\alpha(K_{ij}K^{ij}-\lambda K^{2})+\beta
C_{ij}C^{ij}+\xi R+a_{1}(a_{i}a^{i}))\\\nonumber
a_{i}=\frac{\partial_{i}N}{N}
\end{eqnarray}
The same analysis was performed by this action and by imposing
static gauges for Lapse and the metric functions(with a vanishes
shift function) by  Kiritsis is Spherical symmetry[83]. and we
are going to look for plane symmetric solutions with zero shift of the
form
\begin{eqnarray}
ds^{2}=-N(z)^{2}dt^{2}+dz^{2}+u(z)^{2}(dx^{2}+dy^{2})
\end{eqnarray}
For such  static-gauge the kinetic  terms do not contributes
the equations of motion. We insert this metric ansatz in the
action we have:
\begin{eqnarray}\nonumber
S=\int d^{3}xdt L\\\nonumber L=2\beta N
u'^{2}(\frac{u''}{u}-\frac{u'^{2}}{u^{2}})^{2}+2\xi(u'^{2}+2uu'')+a_{1}u^{2}\frac{N'^{2}}{N}
\end{eqnarray}
\subsection{Solution in the absence of the \emph{Cotton} tensor}
There is only one simple solution in case $\beta=0$. In this case
the contribution from Cotton tensor disappeared. This solution is
written as:
\begin{eqnarray}
N(z)=\sqrt{\frac{\xi}{2a_{1}}}\coth^{-1}(c_{1}\sqrt{\frac{a_{1}}{2\xi}}z)\\
u(z)=c_{2}\frac{N(z)^{1/4}}{\sqrt{N'(z)}}
\end{eqnarray}
\subsection{Constant (spatially) curvature solutions}
We know that constant curvature solutions form are a rich family of
investigations both in GR and f(R)  metric gravity.If we accept
that there is a Cosmological constant term in the universe which
dominated in epoch, thus all the models of the universe
imposed that the field equations govern the spacetime manifolds
must be asymptotically de Sitter. This means that the spacetime
posses a constant Ricci scalar. Also in the context of the string
theory the existence of the constant negative curvature solution
(AdS) is fundamental. Far away from any of this reasons as an
attempt we can suppose that there is a constant spatial Ricci
scalar solution for our field equations in this modified form of
the Horava theory. This family of exact solutions can be obtained
by supposing that for this solution the Ricci scalar $R$ is
constant. Solving the field equation for $u(z)$ we obtain:
\begin{eqnarray}
u(z)=\sqrt{\frac{2A}{R}}\cosh(\theta_{0}+\sqrt{\frac{R}{2A}}z)
\end{eqnarray}
for finding the lapse function $N(z)$ one must solve the following
differential equation
\begin{eqnarray}
\psi'+\psi(\frac{1}{2}\psi+2(\log(u(z)))')=g(z)\\\nonumber
\psi=\frac{N'}{N}\\\nonumber
g(z)=\frac{f(z)}{2a_{1}u(z)^{2}}\\\nonumber f(z)=2\beta
u'(z)(\frac{R}{4}-2(\frac{u'(z)}{u(z)})^{2})
 \end{eqnarray}

This is a standard form of a Riccati equation. There is  not
found a general solution for it, yet for non vanishing values of
$g(z)$. But if we set $\beta=0$ then this differential equation
 solves by simple methods. The final result is:
\begin{eqnarray}
N(z)=c_{2}[2A\tanh^{2}(\sqrt{\frac{R}{2A}}z)+4\sqrt{2AR}c_{1}\tanh(\sqrt{\frac{R}{2A}}z)+4Rc_{1}^{2}]
\end{eqnarray}

 \section{Conclusion}
The Horava-Lifshitz gravity is a power-counting renormalizable
theory, which has anisotropic scale between space and time in
the UV limit, and thus breaks the Lorentz invariance.By applying
this theory to plane symmetric spacetimes, one finds a set of
simple differential equations. This system could be solved
analytically and two family of static Taub like solutions are
obtained. In both families the cosmological constant term is
positive. We mentioned here that we do not have any analytical
continuation on the parameters of model. We observe that in
contraction of the cosmological models based on the Horava
theory, there are no trouble problematic features of model in the
special case $\lambda=1/3,1/4$. So the negative cosmological
constant problem can be solved successfully in the plane
symmetry. In this case, the system has both a  state and a non
static one. The former corresponds to the two families
corresponds to the $\lambda=1/3,1/4$. The first solution
$\lambda=1/3$ is a generalization of the Taub fluid solution in
GR.But the EOS of the fluid in this case is so complicated and
contains Legendre functions. Once the equation of state $w=\frac{
 p}{\rho}$ of a perfect fluid reaches to infinity. If we rewrite the
 field equations in terms of a first order system of autonomous
 equations, this stable critical point coincides with the saddle one. Thus the stable state
is broken. It may be expected  that the equation of state may
change with the energy scale.Actually, if $\lambda=1/3$, this
solution represents the role of an infinite  charge in the
Kehagias-Sfetsos (KS) black hole[85]. We notice that if
$g_{\mu\nu}$ is a solution then $\psi g_{\mu\nu}$ is also a
solution, whenever $\psi$ is constant. From now on it shall be
understood that all line elements can be multiplied by a constant
conformal factor. The non static solution recovers the GR
solution in IR regime. It is homogenous and asymptotically
Robertson-Walker Space-time in comoving coordinations. In spite of
the GR stiff matter solutions, unfortunately there are not many
non-static solutions with only one Killing
vector.We may can state a theorem as: \\
\emph{Theorem: If the metric (4)  satisfies the vacuum field
equations for a Horava type of gravity,  then the perfect fluid
field equations for a stiff fluid  are satisfied by the metric
$(\tilde{4})$ which differs from (3) by the  some simple (but not
so
trivial as the GR) in metric function $\Omega(t,z)$ respectively.}\\
We lead proof of this theorem to another work. The technique
described by the above theorem can be  applied to many solutions:
the class of vacuum metrics (4) belongs to those metrics with an
orthogonally transitive Abelian groups. Moreover, since the metric
function $g_{xx}=g_{yy}=t^2$ does not change under these
generation techniques, the same functions $\sigma$ and $\Omega$
can be used for all vacuum solutions obtained from the same
vacuum seed. A different way of looking at the class of solutions
covered by  this Theorem is for starting from a stiff fluid solution,
performing a transformation to a vacuum solution, applying a
soliton-generation technique and then going back to the stiff fluid:
one then may speak of a solution describing solitons travelling
in the background of a (particular) stiff fluid of higher
symmetry. Of course, since $\sigma$ and $\Omega$ need not be
changed, it suffices to immediately transform only when the vacuum
part of the metric.\\
Also we derived two family of exact solution for the modified
version of Horava theory which recently proposed by Blas et al.It
was shown that there is one solution which it's spatial curvature
is constant.Another solution is obtained in the absence of the
Cotton tensor.

\end{document}